\documentclass[pra,twocolumn,showpacs]{revtex4-1}
\usepackage{amsmath,amscd,amsfonts,amssymb,color}
\usepackage{graphicx,amsfonts,epsf}
\usepackage{epstopdf}
\usepackage{hyperref}
\newcommand{\ie}{\textit{i.e.~}}
\begin{document}
\title{Experimental Classification of 
Entanglement in Arbitrary Three-Qubit States on an NMR
Quantum Information Processor}
\author{Amandeep Singh}
\email{amandeepsingh@iisermohali.ac.in}
\affiliation{Department of Physical Sciences, Indian
Institute of Science Education \& 
Research Mohali, Sector 81 SAS Nagar, 
Manauli PO 140306 Punjab India.}
\author{Harpreet Singh}
\email{harpreetsingh@iisermohali.ac.in}
\affiliation{Department of Physical Sciences, Indian
Institute of Science Education \& 
Research Mohali, Sector 81 SAS Nagar, 
Manauli PO 140306 Punjab India.}
\author{Kavita Dorai}
\email{kavita@iisermohali.ac.in}
\affiliation{Department of Physical Sciences, Indian
Institute of Science Education \& 
Research Mohali, Sector 81 SAS Nagar, 
Manauli PO 140306 Punjab India.}
\author{Arvind}
\email{arvind@iisermohali.ac.in}
\affiliation{Department of Physical Sciences, Indian
Institute of Science Education \& 
Research Mohali, Sector 81 SAS Nagar, 
Manauli PO 140306 Punjab India.}
\begin{abstract}
We undertake experimental detection of the entanglement
present in arbitrary three-qubit pure quantum states on an
NMR quantum information processor. Measurements of only four
observables suffice to experimentally differentiate between
the six classes of states which are inequivalent under
stochastic local operation and classical communication
(SLOCC).  The experimental realization is achieved by
mapping the desired observables onto Pauli $z$-operators of
a single qubit, which is directly amenable to measurement.
The detection scheme is applied to known entangled states as
well as to states randomly generated using a generic scheme
that can construct all possible three-qubit states.  The
results are substantiated via direct full quantum state
tomography as well as via negativity calculations and the
comparison suggests that the protocol is indeed successful
in detecting tripartite entanglement without requiring any
{\it a priori} information about the states.
\end{abstract} 
\maketitle 
\section{Introduction} 
Quantum entanglement plays a fundamental role in quantum
information processing and is a key resource for several
quantum computational and quantum communication
tasks~\cite{horodecki-rmp-09}.  Any experiment aimed at
entanglement generation needs as an integral part, a way to
establish that entanglement has indeed been
generated~\cite{guhne-pr-09}. Therefore, the detection of
entanglement and its characterization is a foundational
problem and is a key focus of research in quantum
information processing~\cite{li-ff-13}.  Entanglement
detection and certification protocols include quantum states
tomography~\cite{thew-pra-02}, entanglement witness
operators~\cite{guhne-jmo-03,
arrazola-pra-12,jungnitsch-prl-11}, of the density operator
under partial transposition~\cite{peres-prl-96,li-sr-17} and
the violation of Bell's inequalities~\cite{peres-pra-97}.

Experimentally, entanglement has been created in various
physical systems including nitrogen-vacancy defect
centers~\citep{neumann-science-08}, trapped-ion quantum
computers~\citep{mandel-nature-03}, superconducting phase
qubits~\citep{neeley-nature-10} nuclear spin
qubits~\citep{dogra-pra-15} and quantum
dots~\citep{gao-nature-12}.  Bound entanglement was created
and detected using three nuclear
spins~\cite{kampermann-pra-10} and there have been several
efforts to create and detect three-qubit entanglement using
NMR~\cite{laflamme-rsta-98,peng-pra-10,rao-ijqi-12,dogra-pra-15-1,xin-pra-18}.
Witness based entanglement detection protocols have been
implemented experimentally in quantum
optics~\citep{bourennane-prl-04} and
NMR~\citep{filgueiras-qip-12}.
Concurrence~\citep{wootters-qic-01} was measured by a single
measurement on twin copies of the quantum state of
photons~\citep{walborn-nature-06} while entanglement of
formation was used as an entanglement quantifier in four
trapped ions~\citep{sackett-nature-00}.  While there have
been various experimental advances to detect entanglement
yet characterizing entanglement experimentally as well as
computationally is a daunting
task~\cite{dur-jpamg-01,altepeter-prl-05,spengler-pra-12,dai-prl-14}.
Therefore it is desirable to invent and implement protocols
to certify the existence of entanglement which are not
intensive on resources.

In the present study we undertake the experimental
characterization of arbitrary three-qubit pure states.  The
three-qubit states can be classified into six inequivalent
classes~\cite{dur-pra-00} under
SLOCC~\citep{bennett-pra-00}.Protocols have been
invented to carry out the classification of three-qubit
states into the SLOCC
classes~\cite{chi-pra-2010,zhao-pra-2013}. A recent proposal
aims to  classify any three-qubit pure entangled state into
these six inequivalent classes by measuring only four
observables~\citep{adhikari-arxiv-17}.  We have previously
constructed a scheme to experimentally realize a canonical
form for general three-qubit states, which we use here to
prepare arbitrary three-qubit states with an unknown amount
of entanglement.  Experimental implementation of the
entanglement detection protocol is such that in a single
shot (using only four experimental settings), we were able
to determine if a state belongs to the $\rm W $ class or to
the ${\rm GHZ}$ class.  We use our own scheme to map the
desired observables onto the $z$-magnetization of one of the
subsystems, making it possible to experimentally measure its
expectation value on NMR systems~\citep{singh-pra-16}.
Mapping of the observables onto Pauli $z$-operators of a
single qubit eases the experimental determination of the
desired expectation value, since the NMR signal is
proportional to the ensemble average of the Pauli
$z$-operator.

We implement  the protocol on known three-qubit entangled
states such as the ${\rm GHZ}$ state and the $W$ state and
also implement it on randomly generated arbitrary
three-qubit states with an unknown amount of entanglement.
Seven representative states  belonging to the six SLOCC
inequivalent classes  as well as twenty random states were
prepared experimentally, with state fidelities ranging
between 89\% to 99\%.  To decide the entanglement class of a
state, the expectation values of four observables were
experimentally measured in the state under investigation.
All the seven representative states (namely, GHZ, W, $\rm
W\overline{W}$, three bi-separable states and a separable
state) were successfully detected within the experimental
error limits.  Using this protocol, the experimentally
randomly generated arbitrary three-qubit states were
correctly identified  as belonging to either the GHZ, the W,
the bi-separable or the separable class of states.  We also
perform full quantum state tomography to directly compute
the observable value. Reconstructed density matrices were
used to calculate the entanglement by computing negativity
in each case, and the results compared well with those of
the current protocol.

The paper is organized as follows: Section~\ref{Theory}
briefly describes the theoretical framework, while the
mapping of the required observables onto single-qubit $z$
magnetization is discussed in Section~\ref{Mapping}.
Section~\ref{NMR Implementation} presents the experimental
implementation of the entanglement characterization protocol
on a three-qubit NMR quantum information processor.
Section~\ref{remarks} contains some concluding remarks.

%%%%%%%%%%%%%%%%%%%%%%%%%%%%%%%%%%%%%%%%%%%%%%%%%%%%
\section{Detecting Tripartite Entanglement} \label{Theory} There are six SLOCC
inequivalent classes of entanglement in three-qubit systems, namely, the GHZ,
W, three different biseparable classes and the separable
class~\cite{dur-pra-00}.  A widely used measure of entanglement is the
$n$-tangle~\cite{wong-pra-01,li-qip-12} and a non-vanishing three-tangle is a
signature of the GHZ entangled class and can hence be used for their detection.
For three parties A, B and C, the three-tangle $\tau$ is defined as
\begin{equation} \label{3tangle}
\tau=C^{2}_{\rm{A(BC)}}-C^{2}_{\rm{AB}}-C^{2}_{\rm{AC}} \end{equation}  with
$C^{}_{\rm{AB}}$ and $C^{}_{\rm{AC}}$ being the concurrence that characterizes
entanglement between A and B, and between A and C respectively;
$C^{}_{\rm{A(BC)}}$ denotes the concurrence between A and the joint state of
the subsystem comprising B and C~\cite{coffman-pra-00}.

The idea of using the three-tangle to investigate entanglement in three-qubit
generic states is particularly interesting and general, as any three-qubit
pure state can be written in the canonical 
form~\cite{acin-prl-01}
\begin{equation}
\label{generic}
\vert\psi\rangle=a_0\vert 000 \rangle + a_1e^{\iota
\theta}\vert 100 \rangle + a_2\vert 101 \rangle + a_3\vert
110 \rangle + a_4\vert 111 \rangle
\end{equation}
where $a_i\geq 0$, $\sum_i a^2_i=1$ and $\theta \in [0,\pi]$, and the class of
states is written in the computational basis $\{ \vert 0 \rangle, \vert 1
\rangle \}$ of the qubits.
The three-tangle for the generic state given in Eq.~\ref{generic} 
turns out to be~\citep{adhikari-arxiv-17} 
\begin{equation}
\tau_{\psi}=4a^2_0a^2_4
\end{equation}
Three-tangle can be
measured experimentally by measuring the expectation value
of the operator
$O=2\sigma^{}_{1x}\sigma^{}_{2x}\sigma^{}_{3x}$,
in the three qubit state $\vert
\psi \rangle$. Here 
$\sigma^{}_{x,y,z}$ are the Pauli matrices and $i=1,2,3$
denotes qubits label and the tensor product symbol 
between the Pauli operators 
has been omitted for brevity.
Since, $\langle \psi \vert O \vert \psi
\rangle^2= \langle O \rangle^{2}_{\psi}=
4\tau_\psi$, a non-zero expectation value of $O$ 
implies that the state under
investigation is in the GHZ class~\citep{dur-pra-00}. 
In
order to further categorize the classes of three-qubit generic
states we need three more observables
$O^{}_1=2\sigma^{}_{1x}\sigma^{}_{2x}\sigma^{}_{3z}$,
$O^{}_2=2\sigma^{}_{1x}\sigma^{}_{2z}\sigma^{}_{3x}$,
$O^{}_3=2\sigma^{}_{1z}\sigma^{}_{2x}\sigma^{}_{3x}$.
Experimentally measuring the expectation values of the
operators $O$, $O^{}_1$, $O^{}_2$ and $O^{}_3$ can reveal
the entanglement class of every three-qubit  
 pure state~\cite{zhao-pra-2013,adhikari-arxiv-17}.
Table~\ref{classification table}
summarizes the classification of the six SLOCC inequivalent
classes of entangled states 
based on the expectation values of the observables
$O$, $O^{}_1$, $O^{}_2$, $O^{}_3$.
%%%%%%%%%%%%%%%%%%%%%%%%%%%%%%%%%%%%%%%
\begin{table}[h]
\caption{\label{classification table}
Decision table for the classification of three qubit pure
entangled states based on the expectation values of
operators $O$, $O^{}_1$, $O^{}_2$ and $O^{}_3$ in state $
\vert \psi \rangle $. Each class in the row is shown with
the expected values of the observables.}
\begin{ruledtabular}
\begin{tabular}{c | r r r r }%{lcdr}
\textrm{Class} &
\textrm{$\langle O \rangle$}&
\textrm{$\langle O^{}_1 \rangle$} &
\textrm{$\langle O^{}_2 \rangle$} &
\textrm{$\langle O^{}_3 \rangle$} \\
\colrule
GHZ    & $\neq 0$ & $*$ & $*$ & $*$  \\
W      & 0 & $\neq 0$ & $\neq 0$ & $\neq 0$ \\ 
BS$_1$ & 0 & 0 & 0 & $\neq 0$ \\ 
BS$_2$ & 0 & 0 & $\neq 0$ & 0 \\ 
BS$_3$ & 0 & $\neq 0$ & 0 & 0 \\ 
Separable & 0 & 0 & 0 & 0 \\ 
\end{tabular}
\begin{flushleft}
$*$ May or may not be zero.
\end{flushleft}
\end{ruledtabular}
\end{table}
%%%%%%%%%%%%%%%%%%%%%%%%%%%%%%%%%%%%%%%% 
The six SLOCC inequivalent classes of three-qubit entangled states are
GHZ, W, BS${}_1$, BS${}_2$, BS${}_3$ and separable. While
GHZ and W classes are well known, BS$_1$ denotes a
biseparable class having B and C subsystems entangled, the
BS$_2$ class has subsystems A and C entangled, while the
BS$_3$ class has subsystems A and B entangled. As has been
summarized in Table~\ref{classification table} a non-zero
value of $\langle O \rangle$ indicates that the state is in
the GHZ class and this expectation value is zero for all
other classes. For the W class of states all $\langle O_j\rangle$
are non-zero except $\langle O\rangle$. For the BS${}_1$ class
only $\langle O_3\rangle$ is non-zero while only $\langle
O_2\rangle$, and $\langle O_1\rangle$  are non-zero for the
classes BS${}_2$ and BS${}_3$, respectively. For separable states
all expectations are zero.

In order to experimentally realize the entanglement
characterization protocol, one has to determine the
expectation values $\langle O \rangle$, $\langle O_1
\rangle$, $\langle O_2 \rangle$ and $\langle O_3 \rangle$
for an experimentally prepared state $\vert \psi \rangle$.
In the next section we will describe our method
to experimentally realize these expectation values based on
subsystem measurement of the Pauli
$z$-operator~\cite{singh-pra-16} and our experimental scheme
for generating arbitrary 3 qubit states~\cite{dogra-pra-15}.
%%%%%%%%%%%%%%%%%%%%%%%%%%%%%%%%%%%%%%%%%%%%%%%%%%%%%%%%%%
\subsection{Mapping Pauli basis operators to single
qubit $z$-operators}
\label{Mapping}
A standard way to determine the expectation value of a
desired observable in an experiment is to decompose the
observable as a linear superposition of the observables
accessible in the experiment~\cite{nielsen-book-02}. This
task becomes particularly accessible while dealing with 
the Pauli basis.

Any observable for a three-qubit system, acting on an 
eight-dimensional Hilbert space can be decomposed as a linear
superposition of 64 basis operators, and the 
Pauli basis is one possible basis for this decomposition.
Let the set of Pauli
basis operators be denoted as $\mathbb{B}=\{ \rm{B}_i; 0\leq
i \leq 63\}$.
For example, $O^{}_2$
has the form $\sigma^{}_{1x}\sigma^{}_{2z}\sigma^{}_{3x}$
and it is the element B$_{29}$ of
the basis set $\mathbb{B}$. 
The four
observables $O$, $O^{}_1$, $O^{}_2$ and $O^{}_3$ are
represented by the elements B$_{21}$, B$_{23}$, B$_{29}$ and
B$_{53}$ respectively of the Pauli basis set $\mathbb{B}$.
Also by this convention the single-qubit $z$-operators for
the first, second and third qubit \ie $\sigma^{}_{1z}$,
$\sigma^{}_{2z}$ and $\sigma^{}_{3z}$ are the elements
B$_{48}$, B$_{12}$ and B$_{3}$ respectively.

%%%%%%%%%%%%%%%%%%%%%%%%%%%%%%%%%%%%%%%
\begin{figure}[h]
\includegraphics[angle=0,scale=1]{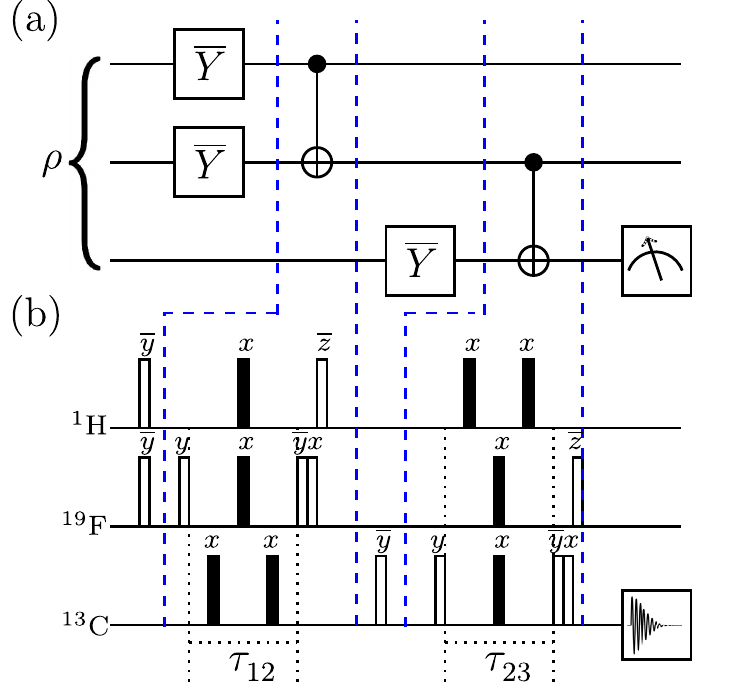}
\caption{(a) Quantum circuit to achieve mapping of the state
$ \rho $ to either of the states $ \rho^{}_{21} $, $
\rho^{}_{23} $, $ \rho^{}_{29} $ or $ \rho^{}_{53} $
followed by measurement of qubit 3 in the computational
basis. (b) NMR pulse sequence of the quantum circuit given
in (a). All the unfilled rectangles denote $ \frac{\pi}{2} $
spin-selective RF pulses while filled rectangles denote $ \pi
$ pulses. Pulse phases are written above the respective pulse 
and a bar over a phase represents negative phase. Delays
are given by $\tau^{}_{ij}=1/(8 J_{ij})$; $i,j$ label
the qubit and $J$ is the coupling constant.}
\label{ckt+seq}
\end{figure}
%%%%%%%%%%%%%%%%%%%%%%%%%%%%%%%%%%%%%%%
Table~\ref{Mapping Table} in Appendix~\ref{App-I} details
the mapping of all 63 Pauli basis operators (excluding
the 8$\otimes$8 identity operator) to the single-qubit Pauli
$z$-operator. This mapping is particularly useful in an
experimental setup where the expectation values of Pauli's
local $z$-operators are easily accessible. In NMR
experiments, the $z$-magnetization of a nuclear spin in a
state is proportional to the expectation value of Pauli
$z$-operator of that spin in the state.

As an example of the mapping given in
Table~\ref{Mapping Table},
the operator $O^{}_2$ has the form
$\sigma^{}_{1x}\sigma^{}_{2z}\sigma^{}_{3x}$ and is the
element B$_{29}$ of basis set $\mathbb{B}$. In order to
determine $\langle O^{}_2 \rangle $ in the state $\rho=\vert
\psi \rangle \langle \psi \vert$, one can map the state
$\rho \rightarrow
\rho^{}_{29}=U^{}_{29}.\rho.U^{\dagger}_{29} $ with
$U^{}_{29}= {\rm CNOT}_{23}.\overline{Y}_3.{\rm
CNOT}_{12}.\overline{Y}_1 $. This is followed by finding
$\langle \sigma^{}_{3z} \rangle $ in the state
$\rho^{}_{29}$. The expectation value $\langle
\sigma^{}_{3z} \rangle $ in the state $\rho^{}_{29}$ is
equivalent to the expectation value of $\langle O^{}_2
\rangle $ in the state $\rho=\vert \psi \rangle \langle \psi
\vert $(Table~\ref{Mapping Table}); the operation ${\rm
CNOT}_{kl}$ is a controlled-NOT gate with $k$ as the control
qubit and $l$ as the target qubit, and $X$, $\overline{X}$,
$Y$ and $\overline{Y}$ represent local $\frac{\pi}{2}$
unitary rotations with phases $x$, $-x$, $y$ and $-y$
respectively.  The subscript on $\pi/2$ local unitary
rotations denotes qubit number. The quantum circuit to
achieve such a mapping is shown in Fig.~\ref{ckt+seq}(a).

It should be noted that while measuring the expectation
values of $O$, $O^{}_1$, $O^{}_2$ or $O^{}_3$, all the
$\overline{Y}$ local rotations may not act in all these four cases.  The mapping given in Table~\ref{Mapping Table} is used to decide which $\overline{Y}$ local rotation in the
circuit~\ref{ckt+seq}(a) will act.  All the basis operators
in set $\mathbb{B}$ can be mapped to single qubit
$z$-operators in a similar fashion.  The mapping given in
Table~\ref{Mapping Table} is not unique and there are
several equivalent mappings which can be worked out as per the experimental requirements.
%%%%%%%%%%%%%%%%%%%%%%%%%%%%%%%%%%%%%%%
\begin{figure}[h]
\includegraphics[angle=0,scale=1.0]{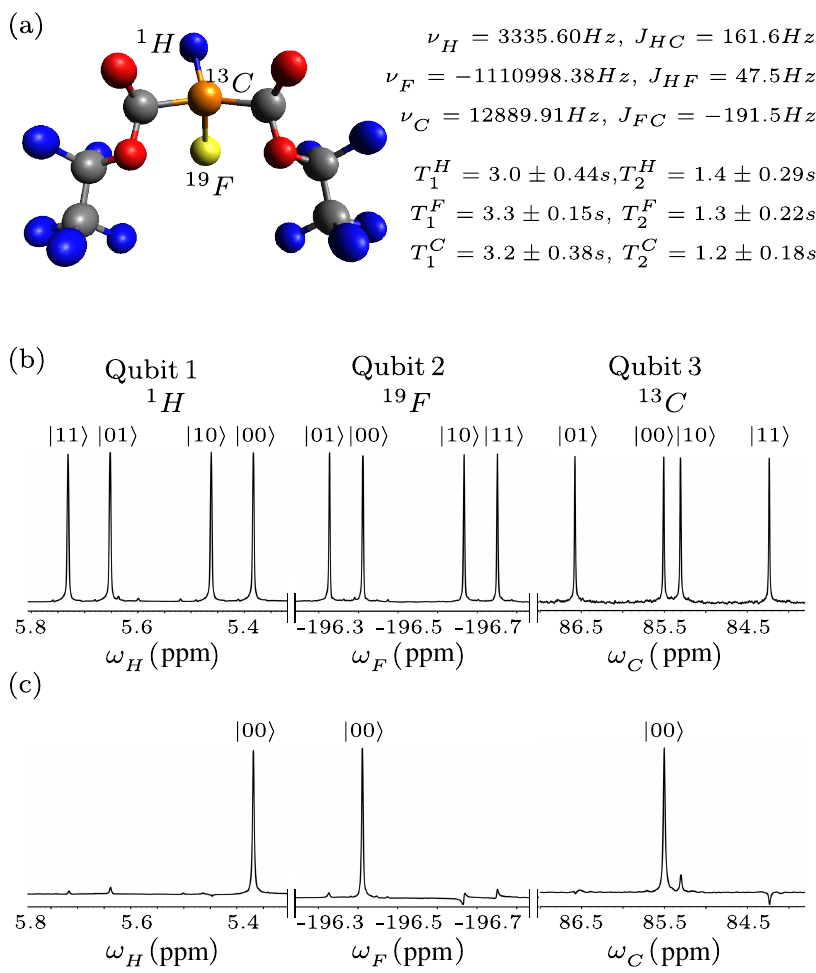}
\caption{(a) Molecular structure of
$^{13}C$-labeled diethyl fluoromalonate and NMR parameters.
NMR spectra of (b) thermal equilibrium state (c) pseudopure
state. Each peak is labeled with the logical state of
the qubit which is passive during the transition. Horizontal scale
represents the chemical shifts in ppm.
}
\label{molecule} 
\end{figure}
%%%%%%%%%%%%%%%%%%%%%%%%%%%%%%%%%%%%%%%%%%%%%%%%%
\section{NMR Implementation of Three Qubit Entanglement
Detection Protocol}
\label{NMR Implementation}
The Hamiltonian~\citep{ernst-book-90} for a three-qubit
system in the rotating frame is given by
\begin{equation}\label{Hamiltonian}
\mathcal{H}= -\sum_{i=1}^{3} \nu_i I_{iz} + \sum_{i
>j,i=1}^{3} J_{ij}I_{iz}I_{jz}
\end{equation}
where the indices $i,j=$ 1,2 or 3 represent the qubit number and
$\nu_i$ is the respective chemical shift in rotating frame,
$J_{ij}$ is the scalar coupling constant and $I_{iz}$ is the
Pauli's $z$-spin angular momentum operator of the
$i^{\rm{th}}$ qubit. To implement the entanglement detection
protocol experimentally, $^{13}$C labeled diethyl
fluoromalonate dissolved in acetone-D6 sample was used.
$^{1}$H, $^{19}$F and $^{13}$C spin-half nuclei were encoded
as qubit 1, qubit 2 and qubit 3 respectively. 
The
system was initialized in the pseudopure (PPS) state \ie $\vert 000
\rangle$ using the spatial averaging~\cite{cory-physD-98,mitra-jmr-07} 
with the density operator being
\begin{equation}
\rho_{000}=\frac{1-\epsilon}{2^3}\mathbb{I}_8 +\epsilon
\vert 000 \rangle \langle 000 \vert 
\end{equation} 
where $\epsilon \sim 10^{-5}$ is the thermal polarization at
room temperature and $ \mathbb{I}_8 $ is the 8 $ \times $ 8
identity operator. The experimentally determined
NMR parameters (chemical shifts,
T$_1$ and T$_2$ relaxation times and 
scalar couplings $\rm{J}_{ij}$) 
as well as the NMR spectra of the 
PPS
state  are shown in Fig.~\ref{molecule}. 
Each spectral transition is labeled with
the logical states of the passive qubits (\ie qubits not
undergoing any transition) 
in the computational basis.
The state fidelity of the experimentally prepared PPS 
(Fig.~\ref{molecule}(c)) was compute to be 
0.98$\pm$0.01 and was calculated
using the fidelity measure
\citep{uhlmann-rpmp-76,jozsa-jmo-94}
\begin{equation}\label{fidelity}
F=\left[Tr\left(
\sqrt{\sqrt{\rho_{{\rm th}}}\rho_{{\rm ex}}
\sqrt{\rho_{{\rm th}}}}\right)\right]^2
\end{equation}
where $\rho_{{\rm th}}$ and $\rho_{{\rm ex}}$ are the theoretically
expected and the experimentally reconstructed density
operators, respectively. Fidelity measure is normalized 
such that $ F\rightarrow 1 $ as
$\rho_{ex}\rightarrow\rho_{th}$. For the experimental
reconstruction of density operator, full quantum state
tomography (QST)\citep{leskowitz-pra-04,singh-pla-16} was
performed using a preparatory pulse set $\left\lbrace III,
XXX, IIY, XYX, YII, XXY, IYY \right\rbrace$, where
$I$ implies ``no operation''. In NMR 
a $\frac{\pi}{2} $ local unitary rotation $X$($Y$) can be
achieved using spin-selective transverse radio frequency
(RF) pulses having phase $x$($y$). 
%%%%%%%%%%%%%%%%%%%%%%%%%%%%%%%%%%%%%%%%%%%%%%%%%%
\begin{figure}[h]
\includegraphics[angle=0,scale=1]{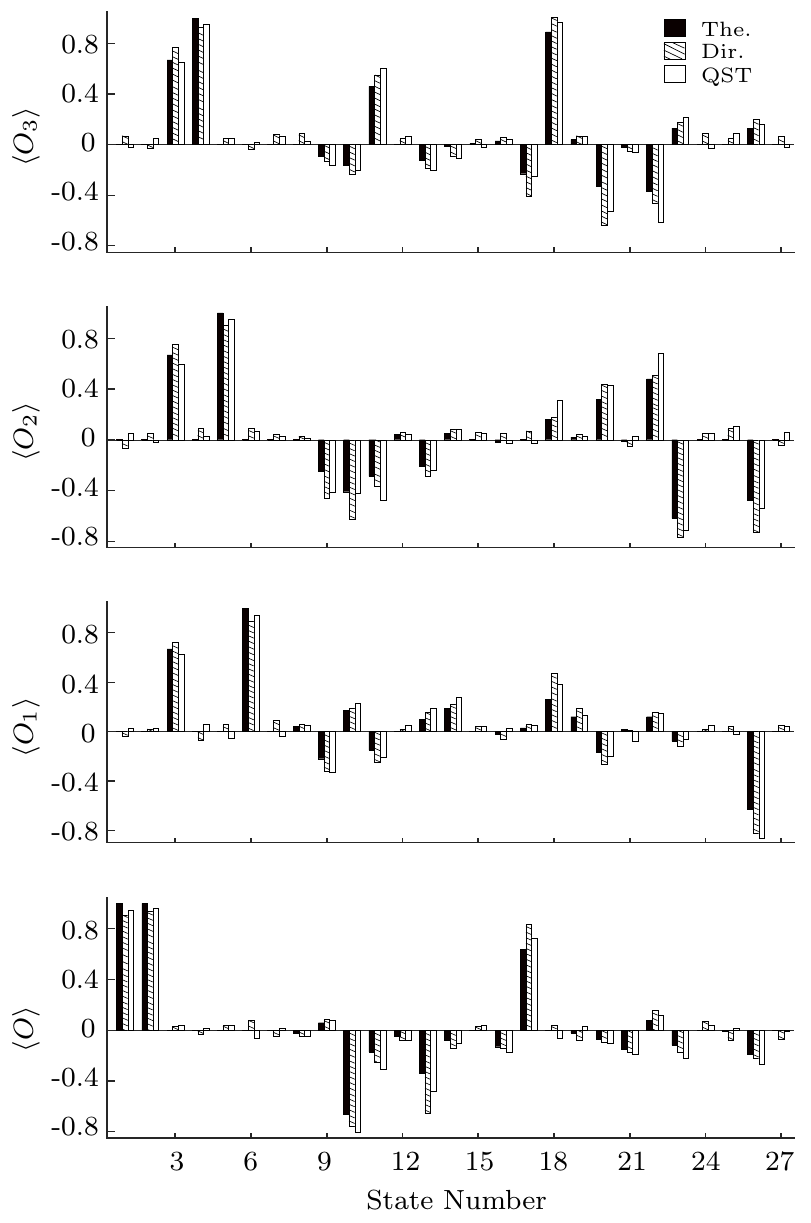}
\caption{Bar plots of the expectation values of the
observables $O$, $O_1$, $O_2$ and $O_3$ for states numbered
from 1-27 (Table~\ref{result table}). The horizontal axes
denote the state number while the vertical axes represent
the values of the respective observable. Black, 
cross-hatched and
unfilled bars represent the theoretical (The.), directly
(Dir.) measured from experiment, and QST-derived expectation
values, respectively.} 
\label{ResultPlot}
\end{figure}
%%%%%%%%%%%%%%%%%%%%%%%%%%%%%%%%%%%%%%%%%%%%%%%%%%%

%%%%%%%%%%%%%%%%%%%%%%%%%%%%%%%%%%%%%%%%%%%%%%%%
\begin{table*} [t]
\caption{\label{result table}
Results of the three qubit entanglement detection protocol
for twenty seven states. Label BS is for biseparable states
while R is for random states. First column depicts the state
label, top row lists the observable (Obs.) while second row
specify if the observable value is theoretical (The.),
direct experimental (Dir.) or from QST.}
\begin{ruledtabular}
%\scriptsize
\begin{tabular}{c | c c c | c c c | c c c | c c c}%{lcdr}

Obs. $\rightarrow$ &  \multicolumn{3}{c}{$\langle O
\rangle$} & \multicolumn{3}{c}{$\langle O^{}_{1} \rangle$} &
\multicolumn{3}{c}{$\langle O^{}_{2} \rangle$} &
\multicolumn{3}{c}{$\langle O^{}_{3} \rangle$} \\

State($F$) $\downarrow$ & The. & Dir. & QST & The. & Dir. &
QST & The. & Dir. & QST & The. & Dir. & QST  \\
\colrule
GHZ(0.95 $\pm$ 0.03) & 1.00 & 0.91 & 0.95 & 0 & -0.04 & 0.03
& 0 & -0.07 & 0.05 & 0 & 0.07 & -0.02 \\
$\rm W\overline{W}$(0.98 $\pm$ 0.01) & 1.00 & 0.94 & 0.96 &
0 & 0.02 & 0.03 & 0 & 0.05 & -0.02 & 0 & -0.03 & 0.05 \\
W(0.96 $\pm$ 0.02) & 0 & 0.05 & 0.04 & 0.67 & 0.60 & 0.62 &
0.67 & 0.61 & 0.69 & 0.67 & 0.59 & 0.63 \\
BS$_1$(0.95 $\pm$ 0.02) & 0 & -0.03 & 0.02 & 0 & -0.07 &
0.06 & 0 & 0.09 & 0.03 & 1.00 & 0.93 & 0.95 \\
BS$_2$(0.96 $\pm$ 0.03) & 0 & 0.04 & 0.04 & 0 & 0.06 & -0.05
& 1.00 & 0.90 & 0.95 & 0 & 0.05 & 0.05 \\
BS$_3$(0.95 $\pm$ 0.04) & 0 & 0.08 & -0.06 & 1.00 & 0.89 &
0.94 & 0 & 0.09 & 0.07 & 0 & -0.04 & 0.02 \\
Sep(0.98 $\pm$ 0.01) & 0 & -0.05 & 0.02 & 0 & 0.09 & -0.04 &
0 & 0.04 & 0.03 & 0 & 0.08 & 0.07 \\
R$_{1}$	(	0.91 $\pm$ 0.02	)	& -0.02	&
-0.05	&	-0.05	&	0.04	&	0.06	&
0.05	&	0.00	&	0.03	&	0.01	&
0.00	&	0.09	&	0.03	\\
R$_{2}$	(	0.94 $\pm$ 0.03	)	&	0.06	&
0.09	&	0.08	&	-0.22	&	-0.32	&
-0.33	&	-0.25	&	-0.46	&	-0.41	&
-0.09	&	-0.13	&	-0.16	\\
R$_{3}$	(	0.93 $\pm$ 0.03	)	&	-0.66	&
-0.76	&	-0.80	&	0.17	&	0.19	&
0.23	&	-0.41	&	-0.63	&	-0.42	&
-0.16	&	-0.23	&	-0.20	\\
R$_{4}$	(	0.91 $\pm$ 0.01	)	&	-0.17	&
-0.25	&	-0.31	&	-0.15	&	-0.25	&
-0.21	&	-0.29	&	-0.37	&	-0.48	&
0.46	&	0.55	&	0.60	\\
R$_{5}$	(	0.94 $\pm$ 0.03	)	&	-0.05	&
-0.08	&	-0.08	&	0.00	&	0.02	&
0.05	&	0.04	&	0.06	&	0.04	&
0.00	&	0.05	&	0.07	\\
R$_{6}$	(	0.90 $\pm$ 0.02	)	&	-0.34	&
-0.65	&	-0.48	&	0.10	&	0.16	&
0.19	&	-0.21	&	-0.29	&	-0.24	&
-0.12	&	-0.19	&	-0.20	\\
R$_{7}$	(	0.93 $\pm$ 0.03	)	&	-0.08	&
-0.14	&	-0.10	&	0.19	&	0.22	&
0.28	&	0.05	&	0.08	&	0.08	&
-0.01	&	-0.09	&	-0.11	\\
R$_{8}$	(	0.94 $\pm$ 0.01	)	&	0.00	&
0.03	&	0.04	&	0.00	&	0.04	&
0.04	&	0.00	&	0.06	&	0.05	&
0.01	&	0.04	&	-0.02	\\
R$_{9}$	(	0.95 $\pm$ 0.02	)	&	-0.13	&
-0.14	&	-0.17	&	-0.02	&	-0.06	&
0.03	&	-0.02	&	0.05	&	-0.03	&
0.03	&	0.06	&	0.04	\\
R$_{10}$	(	0.92 $\pm$ 0.03	)	&	0.64
&	0.84	&	0.73	&	0.03	&	0.06
&	0.05	&	0.00	&	0.07	&
-0.03	&	-0.23	&	-0.41	&	-0.25	\\
R$_{11}$	(	0.93 $\pm$ 0.03	)	&	0.00
&	0.04	&	-0.06	&	0.26	&	0.47
&	0.38	&	0.16	&	0.18	&	0.31
&	0.89	&	1.01	&	0.97	\\
R$_{12}$	(	0.89 $\pm$ 0.02	)	&
-0.02	&	-0.08	&	0.03	&	0.12	&
0.19	&	0.13	&	0.02	&	0.04	&
0.03	&	0.04	&	0.07	&	0.07	\\
R$_{13}$	(	0.92 $\pm$ 0.03	)	&
-0.07	&	-0.09	&	-0.10	&	-0.17	&
-0.26	&	-0.20	&	0.32	&	0.44	&
0.43	&	-0.33	&	-0.64	&	-0.53	\\
R$_{14}$	(	0.94$\pm$ 0.04	)	&
-0.15	&	-0.17	&	-0.19	&	0.02	&
0.01	&	-0.08	&	-0.01	&	-0.05	&
0.03	&	-0.02	&	-0.05	&	-0.06	\\
R$_{15}$	(	0.94 $\pm$ 0.03	)	&	0.08
&	0.16	&	0.12	&	0.12	&	0.16
&	0.15	&	0.48	&	0.51	&	0.68
&	-0.37	&	-0.46	&	-0.61	\\
R$_{16}$	(	0.93 $\pm$ 0.02	)	&
-0.12	&	-0.17	&	-0.22	&	-0.08	&
-0.12	&	-0.06	&	-0.62	&	-0.77	&
-0.71	&	0.13	&	0.18	&	0.22	\\
R$_{17}$	(	0.93 $\pm$ 0.04	)	&	0.00
&	0.07	&	0.04	&	0.00	&	0.02
&	0.05	&	0.00	&	0.05	&	0.05
&	0.00	&	0.09	&	-0.03	\\
R$_{18}$	(	0.90 $\pm$ 0.02	)	&
-0.01	&	-0.08	&	0.02	&	0.00	&
0.04	&	-0.02	&	0.00	&	0.09	&
0.11	&	0.00	&	0.05	&	0.09	\\
R$_{19}$	(	0.94 $\pm$ 0.02	)	&
-0.19	&	-0.22	&	-0.27	&	-0.63	&
-0.82	&	-0.86	&	-0.48	&	-0.73	&
-0.54	&	0.13	&	0.20	&	0.16	\\
R$_{20}$	(	0.93 $\pm$ 0.03	)	&	0.00
&	-0.07	&	-0.01	&	0.00	&	0.05
&	0.04	&	0.00	&	-0.04	&	0.06
&	0.00	&	0.07	&	-0.02	\\

\end{tabular}
\end{ruledtabular}
\end{table*}

%%%%%%%%%%%%%%%%%%%%%%%%%%%%%%%%%%%%%%%%%%%%%%%%%%%%%%%%%%
Experiments were performed 
at room temperature ($293$K) 
on a Bruker Avance III 600-MHz
FT-NMR spectrometer equipped with a QXI probe. 
Local unitary operations were
achieved using highly accurate and calibrated spin
selective transverse RF pulses of suitable amplitude, phase
and duration. Non-local unitary operation were achieved by
free evolution under the system Hamiltonian 
Eq.~\ref{Hamiltonian}, of suitable duration under the desired
scalar coupling with the help of embedded $\pi$ refocusing
pulses. In the current study, the durations of $\frac{\pi}{2}$
pulses for $^{1}$H, $^{19}$F and $^{13}$C were 9.55
$\mu$s at 18.14 W power level, 22.80 $\mu$s at a power level
of 42.27 W and 15.50 $\mu$s at a power level of 
179.47 W, respectively.
\subsection{Measuring Observables by Mapping to Local
$z$-Magnetization}\label{NMR Mapping}
As discussed in Sec.~\ref{Mapping}, the observables
required to differentiate between six inequivalent classes of 
three-qubit pure entangled states can be mapped to the Pauli
$z$-operator of one of the qubits. Further, in
NMR the observed $z$-magnetization of a nuclear spin in a
quantum state is proportional to the expectation value of $
\sigma_{z} $-operator~\citep{ernst-book-90} 
of the spin in that state. The time-domain NMR
signal \ie the free induction decay with appropriate phase
gives Lorentzian peaks when Fourier transformed. These
normalized experimental intensities give an estimate of the
expectation value of $\sigma_{z}$  of the quantum state.

Let $\mathcal{\hat{O}}$ be the observable whose expectation
value is to be measured in a state $ \rho =\vert \psi
\rangle \langle \psi \vert $. Instead of measuring $\langle
\mathcal{\hat{O}}\rangle _{\rho}$, the state $ \rho $ can be
mapped to $ \rho_i $ using $ \rho_i=U_i.\rho.U^{\dagger}_i $
followed by $ z $-magnetization measurement of one of the
qubits. Table~\ref{Mapping Table} lists the explicit forms of
$U_i$ for all the basis elements of the 
Pauli basis set
$\mathbb{B}$. In the present study, the observables of
interest are $O$, $O^{}_1$, $O^{}_2$ and $O^{}_3$  as
described in Sec.~\ref{Mapping} and
Table~\ref{classification table}. 
The quantum circuit to achieve
the required mapping is shown in Fig.~\ref{ckt+seq}(a). The
circuit is designed to map the state $ \rho $
to either of the states $ \rho^{}_{21} $, $ \rho^{}_{23} $,
$ \rho^{}_{29} $ or $ \rho^{}_{53} $ followed by a $\sigma_z$
measurement on the
third qubit in the mapped state. Depending upon the
experimental settings, $ \langle B_3 \rangle $ in the mapped
states is indeed the expectation values of $O$, $O^{}_1$,
$O^{}_2$ or $O^{}_3$ in the initial state $ \rho $.

The NMR pulse sequence to achieve the quantum mapping of
circuit in Fig.~\ref{ckt+seq}(a) is shown in
Fig.~\ref{ckt+seq}(b). The unfilled rectangles represent
$\frac{\pi}{2}$ spin-selective pulses while the filled
rectangles represent $\pi$ pulses. 
Evolution under
chemical shifts has been refocused during all the free
evolution periods 
(denoted by $ \tau_{ij}=\frac{1}{8J_{ij}}$) 
and $\pi$ pulses are embedded in between the free evolution
periods in such a way that the system evolves only under the
desired scalar coupling $J_{ij}$. 

\subsection{Implementing the Entanglement Detection
Protocol}
\label{demo states}
The three-qubit system  was prepared 
in twenty seven different states
in order to experimentally demonstrate the efficacy of  
the entanglement detection protocol.
Seven
representative states were prepared from the six inequivalent
entanglement classes \ie  GHZ (GHZ and $\rm W\overline{W}$
states), W, three bi-separable and a separable class of
states. In addition,
twenty generic states were randomly generated (labeled
as R$_1$, R$_2$, R$_3$,......., R$_{20}$). To prepare the
random states the
MATLAB\textsuperscript{\textregistered}-2016a random number
generator was used. Our recent \citep{dogra-pra-15} 
experimental scheme
was utilized to prepare the generic three-qubit
states. For the details of quantum circuits as well as NMR
pulse sequences used for state preparation see
\citep{dogra-pra-15}. All the prepared states had state fidelities
ranging between 0.89 to 0.99. Each prepared state
$\rho$ was passed through the detection circuit \ref{ckt+seq}(a)
to yield the expectation values of the observables $O$,
$O^{}_1$, $O^{}_2$ and $O^{}_3$ as described in
Sec.~\ref{NMR Mapping}. Further, full QST
\citep{cory-physD-98} was performed  to directly estimate
the expectation value of $O$, $O^{}_1$, $O^{}_2$ and
$O^{}_3$ for all the twenty seven states.

The results of the experimental
implementation of the three-qubit entanglement detection
protocol are tabulated in Table~\ref{result table}. For a
visual representation of the data in Table~\ref{result
table}, bar charts have been plotted  and are shown in
Fig.\ref{ResultPlot}. 
The seven known states were numbered as
1-7 while twenty random states were numbered as 8-27 in
accordance with Table~\ref{result table}. Horizontal axes in
plots of Fig.~\ref{ResultPlot} denote the state number
while vertical axes represent the value of the respective
observable. Black, cross-hatched and unfilled bars represent
theoretical (The.), direct (Dir.) experimental and QST based expectation
values, respectively. To further quantify the entanglement quotient, the
entanglement measure, negativity~\citep{weinstein-pra-10,vidal-pra-02} was also
computed theoretically as well as experimentally in all the cases
(Table~\ref{negativity table}). Experiments were repeated several times for
error estimation and to validate the reproducibility of the experimental
results.
All the seven representative states belonging to the six
inequivalent entanglement classes were detected successfully
within the experimental error limits, as suggested by the experimental results in first seven rows of
Table~\ref{result table} in comparison with
Table~\ref{classification table}.  The errors in the
experimental expectation values reported in the
Table~\ref{result table} were in the range 3.1\%-8.5\%.  The entanglement
detection protocol with only four observables is further supported by
negativity measurements
(Table~\ref{negativity table}).
It is to be noted here that one will never
be able to conclude that the result of an experimental observation
is exactly zero. However it can be established that the result
is non-zero. This has to be kept in mind while interpreting the
experimentally obtained values of
the operators involved via the decision 
Table~\ref{classification table}.
%%%%%%%%%%%%%%%%%%%%%%%%%%%%%%%%%%%%%%%%%%%%%%%%%%
%----------- Negativity Table  -------------------
%%%%%%%%%%%%%%%%%%%%%%%%%%%%%%%%%%%%%%%%%%%%%%%%%%
\begin{table}[t]
\caption{\label{negativity table}
Theoretically calculated and experimentally measured values of negativity.
}
\begin{ruledtabular}
%\scriptsize
\begin{tabular}{c | c  c}%{lcdr}

Negativity $\rightarrow$ &  Theoretical  & Experimental \\

State $\downarrow$ &  &   \\
\colrule
GHZ & 0.5 & 0.46 $\pm$ 0.03 \\
$\rm W\overline{W}$ & 0.37 & 0.35 $\pm$ 0.03 \\
W & 0.47 & 0.41 $\pm$ 0.02 \\
BS$_1$  & 0 & 0.03 $\pm$ 0.02 \\
BS$_2$  & 0 & 0.05 $\pm$ 0.02 \\
BS$_3$  & 0 & 0.03 $\pm$ 0.03 \\
Sep     & 0 & 0.02 $\pm$ 0.01 \\
R$_{1}$ & 0.02&0.04 $\pm$ 0.02\\
R$_{2}$ & 0.16&0.12 $\pm$ 0.04\\
R$_{3}$ & 0.38&0.35 $\pm$ 0.07\\
R$_{4}$ & 0.38&0.34 $\pm$ 0.06\\
R$_{5}$ & 0.03&0.04 $\pm$ 0.02\\
R$_{6}$ & 0.21&0.18 $\pm$ 0.04\\
R$_{7}$ & 0.09&0.08 $\pm$ 0.03\\
R$_{8}$ & 0   &0.02 $\pm$ 0.02\\
R$_{9}$ & 0.07&0.06 $\pm$ 0.03\\
R$_{10}$ &0.38&0.35$\pm$ 0.08\\
R$_{11}$ &0.32&0.28$\pm$ 0.06\\
R$_{12}$ &0.05&0.04$\pm$ 0.02\\
R$_{13}$ &0.18&0.15$\pm$ 0.03\\
R$_{14}$ &0.08&0.07$\pm$ 0.02\\
R$_{15}$ &0.34&0.32$\pm$ 0.06\\
R$_{16}$ &0.30&0.28$\pm$ 0.06\\
R$_{17}$ &0 & 0.03$\pm$ 0.02\\
R$_{18}$ &0 & 0.02$\pm$ 0.02\\
R$_{19}$ &0.39&0.36$\pm$ 0.09\\
R$_{20}$ &0 & 0.02$\pm$ 0.02\\
\end{tabular}
\end{ruledtabular}
\end{table}
%%%%%%%%%%%%%%%%%%%%%%%%%%%%%%%%%%%%%%%%%%%%%%%%%%%%%

The results for the twenty randomly generated generic
states, numbered from 8-27 (R$_1$-R$_{20}$), are
interesting. For instance, states R$_{10}$ and R$_{11}$ have
a negativity of approximately 0.35 which implies that these
states have genuine tripartite entanglement.  On the other
hand the experimental results of current detection protocol
(Table~\ref{result table}) suggest that R$_{10}$ has a
nonzero 3-tangle, which is a signature of the GHZ class. The
states R$_{3}$, R$_{4}$, R$_{6}$, R$_{7}$, R$_{14}$,
R$_{16}$ and R$_{19}$ also belong to the GHZ class as they
all have non-zero 3-tangle as well as finite negativity.  On
the other hand, the state R$_{11}$ has a vanishing 3-tangle
with non-vanishing expectation values of $O_1$, $O_2$ and
$O_3$ which indicates that this state belongs to the W
class. The states  R$_{2}$, R$_{13}$ and R$_{15}$ were also
identified as members of the W class using the detection
protocol.  These results demonstrate the fine-grained state
discrimination power of the entanglement detection protocol
as compared to procedures that rely on QST.  Furthermore,
all vanishing expectation values as well as a near-zero
negativity, in the case of R$_8$ state, imply that it
belongs to the separable class. The randomly generated
states R$_{1}$, R$_{5}$, R$_{17}$, R$_{18}$ and
R$_{20}$ have also been identified as belonging to the
separable class of states. Interestingly, R$_{12}$ has
vanishing values of 3-tangle, negativity, $ \langle O_2
\rangle $ and $ \langle O_3 \rangle $ but has a finite value
of $ \langle O_1 \rangle $, from which one can conclude that
this state belongs to the  bi-separable BS$_3$ class.
%%%%%%%%%%%%%%%%%%%%%%%%%%%%%%%%%%%%%%%%%%%%%%%%%%%%%%%%%%
\section{Concluding Remarks} \label{remarks} We have
implemented a  three-qubit entanglement detection and
classification protocol on an NMR quantum information
processor.  The current protocol is resource efficient as it
requires the measurement of only four observables to detect
the entanglement of unknown three-qubit pure states, in
contrast to the procedures relying on QST, where we need
many more experiments.  The spin ensemble was prepared in a
number of three-qubit states, including standard and
randomly selected states, to test the efficacy of the
entanglement detection scheme.  Experimental results were
further verified and supported with full QST and negativity
measurements.  The protocol was very well able to detect the
entanglement present in the seven representative states
(belonging to the GHZ, W, ${\rm W} {\bar{ \rm W}}$, bi-separable and
separable SLOCC inequivalent classes).  A nonzero
negativity indicates a genuine tripartite
entanglement while a non-vanishing 3-tangle implies that the
state is in GHZ class, and for the randomly generated
states, the protocol was able to classify the R$_{3}$,
R$_{4}$, R$_{6}$, R$_{7}$, R$_{10}$, R$_{14}$, R$_{16}$ and
R$_{19}$ states as belonging to the GHZ class.  Although the
randomly generated R$_{11}$ state has a non-zero negativity,
it has a vanishing 3-tangle, which implies that state
belongs to W class (which is further supported by non-zero
values of the expectation values $O^{}_1$, $O^{}_2$ and
$O^{}_3$).  The states R$_{2}$, R$_{13}$ and R$_{15}$ were
also found to belong to the W class.  Vanishing expectation
values for all the four observables as well as vanishing
negativity values indicate that the randomly generated
states R$_{1}$, R$_{5}$, R$_{8}$, R$_{17}$, R$_{18}$ and
R$_{20}$ belong to the separable class, while the state
R$_{12}$ was correctly identified as belonging to the
BS$_{3}$ class.

With these encouraging experimental results, it would be
interesting to  extend the scheme to mixed states of three
qubits, to a larger number of qubits, and to multipartite 
entanglement detection in higher-dimensional qudit
systems. Results in these directions will be
taken up elsewhere.
Experimentally classifying
entanglement in arbitrary multipartite entangled states is a
challenging venture and our scheme is a step forward in this
direction.
%%%%%%%%%%%%%%%%%%%%%%%%%%%%%%%%%%%%%%%%%%%%%%%%%%%%%%%%%%%%
\begin{acknowledgments}
All the experiments were performed on a Bruker Avance-III
600 MHz FT-NMR spectrometer at the NMR Research Facility of
IISER Mohali. Arvind acknowledges funding from DST India
under Grant No. EMR/2014/000297. K.D. acknowledges funding
from DST India under Grant No. EMR/2015/000556.
\end{acknowledgments}
%\bibliography{3Q-Ent-Det-BIB}
%merlin.mbs 2010-03-15 4.21a (PWD, AO, DPC)
%Control: key (0)
%Control: author (8) initials jnrlst
%Control: editor formatted (1) identically to author
%Control: production of article title (-1) disabled
%Control: page (0) single
%Control: year (1) truncated
%Control: production of eprint (0) enabled
%
\appendix
\section{Mapping Table} 
\label{App-I}
Table~\ref{Mapping Table} lists the explicit form of the unitary
operators, $U_i$, used in the mapping of observables discussed in
Sec.~\ref{Mapping} and \ref{NMR Mapping}.

\begin{table*}[b]
\caption{\label{Mapping Table}
All sixty three product operators, for a three spin (half)
system, mapped to the Pauli $z$-operators (of either spin
1, spin 2 or spin 3) by mapping initial state $ \rho
\rightarrow \rho_i=U_i.\rho.U_i^{\dagger} $.}
\begin{ruledtabular}
%\scriptsize
\begin{tabular}{c c c c}%{lcdr}
\textrm{Observable}&
\textrm{Initial State Mapped via}&
\textrm{Observable}&
\textrm{Initial State Mapped via}\\
\colrule
$\langle B_{1} \rangle$ = Tr[$\rho_{1}.I_{3z}$] & $
U_{1}=\overline{Y}_3 $ & $\langle B_{33} \rangle$ =
Tr[$\rho_{33}.I_{3z}$] & $
U_{33}={\rm CNOT}_{13}.\overline{Y}_3.X_1 $ \\

$\langle B_{2} \rangle$ = Tr[$\rho_{2}.I_{3z}$] & $
U_{2}=X_3 $ & $\langle B_{34} \rangle$ =
Tr[$\rho_{34}.I_{3z}$] & $ U_{34}={\rm CNOT}_{13}.X_3.X_1 $ \\ 

$\langle B_{3} \rangle$ = Tr[$\rho_{3}.I_{3z}$] & $
U_{3}=\mathbb{I}_8 $ & $\langle B_{35} \rangle$ =
Tr[$\rho_{35}.I_{3z}$] & $ U_{35}={\rm CNOT}_{13}.X_1 $ \\

$\langle B_{4} \rangle$ = Tr[$\rho_{4}.I_{2z}$] & $
U_{4}=\overline{Y}_2 $ & $\langle B_{36} \rangle$ =
Tr[$\rho_{36}.I_{2z}$] & $
U_{36}={\rm CNOT}_{12}.\overline{Y}_2.X_1 $ \\

$\langle B_{5} \rangle$ = Tr[$\rho_{5}.I_{3z}$] & $
U_{5}={\rm CNOT}_{23}.\overline{Y}_3.\overline{Y}_2 $ & $\langle
B_{37} \rangle$ = Tr[$\rho_{37}.I_{3z}$] & $
U_{37}={\rm CNOT}_{23}.\overline{Y}_3.{\rm CNOT}_{12}.\overline{Y}_2.X_1
$ \\

$\langle B_{6} \rangle$ = Tr[$\rho_{6}.I_{3z}$] & $
U_{6}={\rm CNOT}_{23}.X_3.\overline{Y}_2 $ & $\langle B_{38}
\rangle$ = Tr[$\rho_{38}.I_{3z}$] & $
U_{38}={\rm CNOT}_{23}.X_3.{\rm CNOT}_{12}.\overline{Y}_2.X_1 $ \\

$\langle B_{7} \rangle$ = Tr[$\rho_{7}.I_{3z}$] & $
U_{7}={\rm CNOT}_{23}.\overline{Y}_2 $ & $\langle B_{39} \rangle$
= Tr[$\rho_{39}.I_{3z}$] & $
U_{39}={\rm CNOT}_{23}.{\rm CNOT}_{12}.\overline{Y}_2.X_1 $ \\

$\langle B_{8} \rangle$ = Tr[$\rho_{8}.I_{2z}$] & $
U_{8}=X_2 $ & $\langle B_{40} \rangle$ =
Tr[$\rho_{40}.I_{2z}$] & $ U_{40}={\rm CNOT}_{12}.X_2.X_1 $ \\
%-----------------------------------------------------------------------------
$\langle B_{9} \rangle$ = Tr[$\rho_{9}.I_{3z}$] & $
U_{9}={\rm CNOT}_{23}.\overline{Y}_3.X_2 $ & $\langle B_{41}
\rangle$ = Tr[$\rho_{41}.I_{3z}$] & $
U_{41}={\rm CNOT}_{23}.\overline{Y}_3.{\rm CNOT}_{12}.X_2.X_1 $ \\

$\langle B_{10} \rangle$ = Tr[$\rho_{10}.I_{3z}$] & $
U_{10}={\rm CNOT}_{23}.X_3.X_2 $ & $\langle B_{42} \rangle$ =
Tr[$\rho_{42}.I_{3z}$] & $
U_{42}={\rm CNOT}_{23}.X_3.{\rm CNOT}_{12}.X_2.X_1 $ \\

$\langle B_{11} \rangle$ = Tr[$\rho_{11}.I_{3z}$] & $
U_{11}={\rm CNOT}_{23}.X_2 $ & $\langle B_{43} \rangle$ =
Tr[$\rho_{43}.I_{3z}$] & $
U_{43}={\rm CNOT}_{23}.{\rm CNOT}_{12}.X_2.X_1 $ \\

$\langle B_{12} \rangle$ = Tr[$\rho_{12}.I_{3z}$] & $
U_{12}=\mathbb{I}_8 $ & $\langle B_{44} \rangle$ =
Tr[$\rho_{44}.I_{2z}$] & $ U_{44}={\rm CNOT}_{12}.X_1 $ \\

$\langle B_{13} \rangle$ = Tr[$\rho_{13}.I_{3z}$] & $
U_{13}={\rm CNOT}_{23}.\overline{Y}_3 $ & $\langle B_{45} \rangle$
= Tr[$\rho_{45}.I_{3z}$] & $
U_{45}={\rm CNOT}_{23}.\overline{Y}_3.{\rm CNOT}_{12}.X_1 $ \\

$\langle B_{14} \rangle$ = Tr[$\rho_{14}.I_{3z}$] & $
U_{14}={\rm CNOT}_{23}.X_3 $ & $\langle B_{46} \rangle$ =
Tr[$\rho_{46}.I_{3z}$] & $
U_{46}={\rm CNOT}_{23}.X_3.{\rm CNOT}_{12}.X_1 $ \\

$\langle B_{15} \rangle$ = Tr[$\rho_{15}.I_{3z}$] & $
U_{15}={\rm CNOT}_{23} $ & $\langle B_{47} \rangle$ =
Tr[$\rho_{47}.I_{3z}$] & $ U_{47}={\rm CNOT}_{23}.{\rm CNOT}_{12}.X_1 $
\\

$\langle B_{16} \rangle$ = Tr[$\rho_{16}.I_{1z}$] & $
U_{16}=X_1 $ & $\langle B_{48} \rangle$ =
Tr[$\rho_{48}.I_{1z}$] & $ U_{48}=\mathbb{I}_8 $ \\
%--------------------------------------------------------------------------------
$\langle B_{17} \rangle$ = Tr[$\rho_{17}.I_{3z}$] & $
U_{17}={\rm CNOT}_{13}.\overline{Y}_3.\overline{Y}_1 $ & $\langle
B_{49} \rangle$ = Tr[$\rho_{49}.I_{3z}$] & $
U_{49}={\rm CNOT}_{13}.\overline{Y}_3 $ \\

$\langle B_{18} \rangle$ = Tr[$\rho_{18}.I_{3z}$] & $
U_{18}={\rm CNOT}_{13}.X_3.\overline{Y}_1 $ & $\langle B_{50}
\rangle$ = Tr[$\rho_{50}.I_{3z}$] & $ U_{50}={\rm CNOT}_{13}.X_3 $
\\

$\langle B_{19} \rangle$ = Tr[$\rho_{19}.I_{3z}$] & $
U_{19}={\rm CNOT}_{13}.\overline{Y}_1 $ & $\langle B_{51} \rangle$
= Tr[$\rho_{51}.I_{3z}$] & $ U_{51}={\rm CNOT}_{13} $ \\

$\langle B_{20} \rangle$ = Tr[$\rho_{20}.I_{2z}$] & $
U_{20}={\rm CNOT}_{12}.\overline{Y}_2.\overline{Y}_1 $ & $\langle
B_{52} \rangle$ = Tr[$\rho_{52}.I_{2z}$] & $
U_{52}={\rm CNOT}_{12}.\overline{Y}_2 $ \\

$\langle B_{21} \rangle$ = Tr[$\rho_{21}.I_{3z}$] & $
U_{21}={\rm CNOT}_{23}.\overline{Y}_3.{\rm CNOT}_{12}.\overline{Y}_2.\overline{Y}_1
$ & $\langle B_{53} \rangle$ = Tr[$\rho_{53}.I_{3z}$] & $
U_{53}={\rm CNOT}_{23}.\overline{Y}_3.{\rm CNOT}_{12}.\overline{Y}_2 $
\\

$\langle B_{22} \rangle$ = Tr[$\rho_{22}.I_{3z}$] & $
U_{22}={\rm CNOT}_{23}.X_3.{\rm CNOT}_{12}.\overline{Y}_2.\overline{Y}_1
$ & $\langle B_{54} \rangle$ = Tr[$\rho_{54}.I_{3z}$] & $
U_{54}={\rm CNOT}_{23}.X_3.{\rm CNOT}_{12}.\overline{Y}_2 $ \\

$\langle B_{23} \rangle$ = Tr[$\rho_{23}.I_{3z}$] & $
U_{23}={\rm CNOT}_{23}.{\rm CNOT}_{12}.\overline{Y}_2.\overline{Y}_1 $ &
$\langle B_{55} \rangle$ = Tr[$\rho_{55}.I_{3z}$] & $
U_{55}={\rm CNOT}_{23}.{\rm CNOT}_{12}.\overline{Y}_2 $ \\

$\langle B_{24} \rangle$ = Tr[$\rho_{24}.I_{2z}$] & $
U_{24}={\rm CNOT}_{12}.X_2.\overline{Y}_1 $ & $\langle B_{56}
\rangle$ = Tr[$\rho_{56}.I_{2z}$] & $ U_{56}={\rm CNOT}_{12}.X_2 $
\\
%-----------------------------------------------------------------------------
$\langle B_{25} \rangle$ = Tr[$\rho_{25}.I_{3z}$] & $
U_{25}={\rm CNOT}_{23}.\overline{Y}_3.{\rm CNOT}_{12}.X_2.\overline{Y}_1
$ & $\langle B_{57} \rangle$ = Tr[$\rho_{57}.I_{3z}$] & $
U_{57}={\rm CNOT}_{23}.\overline{Y}_3.{\rm CNOT}_{12}.X_2 $ \\

$\langle B_{26} \rangle$ = Tr[$\rho_{26}.I_{3z}$] & $
U_{26}={\rm CNOT}_{23}.X_3.{\rm CNOT}_{12}.X_2.\overline{Y}_1 $ &
$\langle B_{58} \rangle$ = Tr[$\rho_{58}.I_{3z}$] & $
U_{58}={\rm CNOT}_{23}.X_3.{\rm CNOT}_{12}.X_2 $ \\

$\langle B_{27} \rangle$ = Tr[$\rho_{27}.I_{3z}$] & $
U_{27}={\rm CNOT}_{23}.{\rm CNOT}_{12}.X_2.\overline{Y}_1 $ & $\langle
B_{59} \rangle$ = Tr[$\rho_{59}.I_{3z}$] & $
U_{59}={\rm CNOT}_{23}.{\rm CNOT}_{12}.X_2 $ \\

$\langle B_{28} \rangle$ = Tr[$\rho_{28}.I_{2z}$] & $
U_{28}={\rm CNOT}_{12}.\overline{Y}_1 $ & $\langle B_{60} \rangle$
= Tr[$\rho_{60}.I_{2z}$] & $ U_{60}={\rm CNOT}_{12} $ \\

$\langle B_{29} \rangle$ = Tr[$\rho_{29}.I_{3z}$] & $
U_{29}={\rm CNOT}_{23}.\overline{Y}_3.{\rm CNOT}_{12}.\overline{Y}_1 $ &
$\langle B_{61} \rangle$ = Tr[$\rho_{61}.I_{3z}$] & $
U_{61}={\rm CNOT}_{23}.\overline{Y}_3.{\rm CNOT}_{12} $ \\

$\langle B_{30} \rangle$ = Tr[$\rho_{30}.I_{3z}$] & $
U_{30}={\rm CNOT}_{23}.X_3.{\rm CNOT}_{12}.\overline{Y}_1 $ & $\langle
B_{62} \rangle$ = Tr[$\rho_{62}.I_{3z}$] & $
U_{62}={\rm CNOT}_{23}.X_3.{\rm CNOT}_{12} $ \\

$\langle B_{31} \rangle$ = Tr[$\rho_{31}.I_{3z}$] & $
U_{31}={\rm CNOT}_{12}.{\rm CNOT}_{23}.\overline{Y}_1 $ & $\langle
B_{63} \rangle$ = Tr[$\rho_{63}.I_{3z}$] & $
U_{63}={\rm CNOT}_{23}.{\rm CNOT}_{12} $ \\

$\langle B_{32} \rangle$ = Tr[$\rho_{32}.I_{1z}$] & $
U_{32}=X_1 $ &  &  \\

\end{tabular}
\end{ruledtabular}
\end{table*}
%%%%%%%%%%%%%%%%%%%%%%%%%%%%%%%%%%%%%%%%%%%%%%%%%%%%%%%
\end{document}